# Frequent Item-set Mining without Ubiquitous Items

Ran M. Bittmann
Machine Learning Center
SAP, Israel
ran.bittmann@sap.com

Philippe Nemery
SAP, Belgium
philippe.nemery@sap.com

Xingtian Shi
SAP, China
xingtian.shi@sap.com

Michael Kemelmakher
Machine Learning Center
SAP, Israel
michael.kemelmakher@sap.com

Mengjiao Wang
SAP, China
mengjiao.wang@sap.com

# Abstract

*Frequent Item-set Mining (FIM), sometimes called Market Basket Analysis (MBA) or Association Rule Learning (ARL), are Machine Learning (ML) methods for creating rules from datasets of transactions of items. Most methods identify items likely to appear together in a transaction based on the support (i.e. a minimum number of relative co-occurrence of the items) for that hypothesis. Although this is a good indicator to measure the relevance of the assumption that these items are likely to appear together, the phenomenon of very frequent items, referred to as ubiquitous items, is not addressed in most algorithms. Ubiquitous items have the same entropy as infrequent items, and not contributing significantly to the knowledge. On the other hand, they have strong effect on the performance of the algorithms and sometimes preventing the convergence of the FIM algorithms and thus the provision of meaningful results. This paper discusses the phenomenon of ubiquitous items and demonstrates how ignoring these has a dramatic effect on the computation performances but with a low and controlled effect on the significance of the results.*

**Keywords**: Machine learning, Frequent item-set mining

# 1. Introduction

Frequent Item-set Mining (FIM), sometimes called Market Basket Analysis (MBA) or Association Rule Learning (ARL), are Machine Learning (ML) methods for creating rules from datasets of observed transactions of items. Generally, the generated rules have the following format:

$$< items\ known\ to\ exist\ in\ a\ basket > \Rightarrow\ < items\ likely\ to\ exist\ in\ the\ basket\ as\ well >$$

where a basket is a collection of items. A transaction is also a basket but it contains all the items accessed, purchased, downloaded … simultaneously in the past. Therefore, a transaction contains many baskets. Typical examples of transactions are shopping baskets, files downloads, visited webpages, etc.

In the rest of the paper, we will adopt the following notation. Let $\mathcal{J} = \{i_1, i_2,\ldots,i_n,\}$ be a set of items. Let $\mathcal{D}$ be a set of transactions where a transaction $\mathcal{T}_l$ is a set of items such that $\mathcal{T}_l \subseteq \mathcal{J}$. $\mathcal{X}$ is a set of items such that $\mathcal{X} \subseteq \mathcal{J}$. We say that $\mathcal{T}_l$ contains $\mathcal{X}$ if $\mathcal{X} \subseteq \mathcal{T}_l$.

Let $n$ be the number of transactions in $\mathcal{D}$, $|\mathcal{D}| = n$, and let $m$ be the number of transactions in $\mathcal{D}$ that contain $\mathcal{X}$. As an example, let us consider the following transactions of customers of a supermarket:

- $\mathcal{T}_1$: bread, milk, sugar
- $\mathcal{T}_2$: coffee, eggs, milk, sugar
- $\mathcal{T}_3$: meat, salt

A basket is for instance {*eggs, sugar*}. This basket is contained in $\mathcal{T}_2$. Basket {*eggs, salt*} on the other hand does not appear in any transaction whereas basket {*coffee, eggs, milk, sugar*} is a basket identical to transaction $\mathcal{T}_2$. The items observed in the transactions are: $\mathcal{J} = \{$*meat, bread, milk, sugar, coffee, eggs, salt*$\}$. The FIM methods are ML algorithms since we feed the computer with historical transactions processed by computer programs / algorithms in order to "learn" from these and to summarize the relations between the items and "suggest" rules to define these relations.

FIM algorithms have been extensively researched recently and let us cite amongst others the following works (Agrawal, et al., 1994) (Agrawal, et al., 1993) (Borgelt, 2003) (Schmidt-Thieme, 2004) (Han, et al., 2000) (Özkural, 2012). Recent works include probabilistic approaches (Wang, et al., 2010) (Tong, et al., 2012), optimization of the features used for the item-set mining (Thirumuruganathan, et al., 2014), application of identifying similar items in databases (Kenig, et al., 2013), privacy issues when using transaction datasets (Zeng, et al., 2012) (Li, et al., 2012).

As pointed out by (Yang, 2004), the number of possible relations between all the different items grows exponentially with the number of items. This makes it usually impossible to use brute force methods to compute and extract rules as presented in section 3.

To address this main obstacle, we propose the introduction of a new parameter, the *ubiquitousness* parameter, which leads to the decrease of the computation time while maintaining a similar quality of the model. Analogously to the filtering of infrequent items (based on the support, etc.), the underlying idea is to avoid the consideration of ubiquitous (very frequent) items in the creation of the rules. As we will see this reduces complexity of the calculations significantly and allows analysis of datasets that previously could not have been analyzed effectively.

The paper is organized as follows. After an overview of the state of the art (An Overview of the State of the Art section 2) we introduce the ubiquitousness parameter (Removing Ubiquitous Items section 3) and its effect on the performance (Performance Analysis section 4), we illustrate the computation gains and analyze the effect on the quality of the obtained rules with experimentations (Experiment Results section 5). The paper finishes with our conclusions and future research directions (Conclusion and Future Work section 6).

## 2. An Overview of the State of the Art

As presented in section 1 FIM are ML methods to create rules based on previous transactions, i.e. set of items that were accessed together. The principle of many of the algorithms is analogous and uses similar parameters:

***Support*** – The (relative) occurrence of a basket in the transactions is used to measure its significance. The *support* can be expressed in total occurrences or by means of a fraction compared to the total number of transactions. A basket has to appear minimum times to be considered as 'relevant' enough. E.g. if we have 100 transactions and we set the support to 0.3 (or 30%) than in order to be considered, a basket needs to appear at least in 30 of the transactions. By "appear" we do not require that the transaction is identical to the basket, but that the basket is contained in the transaction. We call a basket that has enough support a *frequent* basket. With the notation introduced in section 1, we say that $\mathcal{X}$ has support $s$ if $\frac{m}{n} \geq s$.

***Confidence*** – Given a basket, a rule predicting the right hand side (rhs) items given the left hand side (lhs) items, needs the items in the rhs appear a minimum number of times in all the transactions containing the lhs items. The confidence threshold defines the minimum ratio of the transactions containing the items both in the lhs and rhs and the transactions containing the items in the lhs. For instance, let us assume that the basket $\{bread, eggs\}$ is frequent. The rule $bread \Rightarrow eggs$ will said to have a confidence level of 0.7 (70%) if there are 10 transactions containing *bread* and amongst those at least 7 transactions containing both *bread* and *eggs*. Let us remark that this is not commutative such that the rule $bread \Rightarrow eggs$ does not necessary imply the rule $eggs \Rightarrow bread$ holds as well. Furthermore, if $\mathcal{X}$ and $\mathcal{Y}$ are sets of items such that $\mathcal{X} \subset \mathcal{J}$ and $\mathcal{Y} \subset \mathcal{J}$ and $\mathcal{X} \cap \mathcal{Y} = \emptyset$. Let $p$ be the number of transactions in $\mathcal{D}$ that contain $\mathcal{X}$ and let $\ell$ be the number of transactions in $\mathcal{D}$ that contain $\mathcal{X} \cup \mathcal{Y}$. We say that an association rule $\mathcal{X} \Longrightarrow \mathcal{Y}$ holds with confidence $c$ if $\frac{\ell}{p} \geq c$.

The FIM algorithms usually work in two phases:
1. Find all the *frequent* baskets (i.e. baskets with a support equal or higher than a predefined threshold)
2. Amongst the baskets found in phase 1 find the rules that satisfy the *confidence* threshold

3.

The whole process is given in the following pseudo-code:

```
1)    get support s
2)    get confidence c
3)    set rule set 𝒜 = ∅
4)    find ℬ all the item sets 𝒳.support ≥ s
5)    foreach set 𝒳 ∈ ℬ
6)       break 𝒳 into 𝒬 all possible set pairs ℒ, ℛ so that:
7)          𝒳 = ℒ ∪ ℛ and ℒ ≠ ∅ and ℛ ≠ ∅
8)       foreach 𝒬
9)          if 𝒬.confidence ≥ c than
10)            add ℒ ⟹ ℛ to 𝒜
11)         endif
12)      endfor
13)   endfor
14)   return 𝒜
```

Unfortunately, the number of combinations of items, i.e. baskets to be investigated, grows exponentially with the total number of items in all the transactions. In particular, if there are $I$ items then each item may or may not appear in a basket, which results $2^I$ possible baskets. Since empty baskets or single-item based baskets are irrelevant the exact number of possible baskets is $2^I - I - 1$, still exponential making it impractical to calculate rules using brute force. Nevertheless, the Downward Closure Lemma reduces the search space.

The Downward Closure Lemma says that if a basket is *frequent*, than all the baskets contained in it are also *frequent*, or if a basket is *not frequent* (*infrequent*), than there is no basket containing it that can be *frequent*.

This property allows us to prune our search for frequent baskets and avoid looking for baskets containing a combination of items defined as *infrequent*.

Many of the FIM algorithms rely on this property. Let us cite for instance Apriori, fp-growth, Éclat, and LCM (Schmidt-Thieme, 2004) (Borgelt, 2003).

## 3. Removing Ubiquitous Items

As identified in section 2 the exponential growth of baskets to test is a significant problem that is overcome in many cases by applying the Downward Closure Lemma. However this approach does not always work. If there are items that appear in (too) many transactions, then the result will consist of many baskets due to these very frequent items combined with all other baskets. These items are so-called *ubiquitous* items. Borrowing from information theory, if we say that items with low support have a low *Entropy*: these items do not contribute to the *learning* of relevant rules from historical transactions (Cover, et al., 1991). If we consider the item's support as the equivalent of the items probability to appear in a transaction, based on its history, then very frequent items have the same low entropy. For example, assuming the support is a good estimation for the probability of an item, the entropy of an item with support 0.2 is the same as the entropy for an item with support 1 - 0.2 = 0.8.

*Equation 1*

$$H(X) = -\sum_{x \in X} p(x) \cdot \log_2 p(x)$$

Where $H(X)$ is the entropy (Cover, et al., 1991), and $X$ can have the values 1 the item appears and 0 the item does not appear. In our example:

$$H(X) = -0.2 \cdot \log_2 0.2 - 0.8 \cdot \log_2 0.8 = -0.8 \cdot \log_2 0.8 - 0.2 \cdot \log_2 0.2$$

For instance, let us consider again the case of the supermarket. If only one transaction contained the items of a new *fish flavored ice cream* occurring with *bread*, then we cannot deduct that if one buys the *fish flavored ice cream,* one will also buy *bread.* This is simply because there is not enough support for the item *fish flavored ice cream*. In the same manner if nearly all transactions contain a *paper bag*, there exists a *frequent* basket for all the *frequent* items and a *paper bag.* This would lead to the presence of a paper bag in any lhs. Therefore, *ubiquitous* items do not contribute to our knowledge.

However, ubiquitous items do lead to an increase in complexity of the algorithm, and if they are present, they jeopardize the ability to converge and get any (meaningful) set of rules out from the dataset. Therefore, we suggest adding a new threshold to the FIM algorithms called *ubiquitousness parameter,* noted *u*.

The *ubiquitousness* threshold is similar to the *support* threshold but, it filters out the items with a frequency higher than the threshold (rather than the ones with a low frequency). For instance, if there are 100 transactions and the *ubiquitousness* threshold is set to 0.7 (70%), then items that appear in more than 70 transactions are filtered out, and the *frequent* baskets, now called *relevant* baskets may not contain the *ubiquitous* items.

Let us remark that this requires very little modification to the (existing) algorithms themselves, since the items are counted in the first step and the filtering out needs to be performed only once.

The new FIM algorithms would look as follows:

1. Remove all *ubiquitous* items (items whose support is above the ubiquitousness threshold)
2. Find all the *frequent* baskets (i.e. baskets with a support higher than a predefined threshold)
3. Amongst the baskets found in phase 1 find the rules that satisfy the *confidence* threshold

I.e.:

```
1)     get support s
2)     get confidence c
3)     get ubiquitousness u
4)     set rule set A = ∅
5)     remove all items where i.support > u
6)     find B all the item sets where X.support ≥ s
7)     foreach set X ∈ B
8)        break X into Q all possible set pairs L,R so that:
9)           X = L ∪ R and L ≠ ∅ and R ≠ ∅
10)       foreach Q
11)          if Q.confidence ≥ c than
12)             add L ⟹ R to A
13)          endif
14)       endfor
15)    endfor
16)    return A
```

The additions to the original algorithm are underlined.

# 4. Performance Analysis

As discussed in section 2 the number of combination grows exponentially with the number of items. This is overcome by pruning baskets with *infrequent* items. But *frequent* items cannot be trimmed.

Let us consider extreme case where the *ubiquitous* items have a frequency of 100%, then:
- These items have no contribution but appear in each transaction. We cannot make any assumption if they are in lhs of a rule, and they are always in rhs of a rule (if they are not in the lhs of that rule).
- The ubiquitous items' contribution to the complexity of the algorithm is exponential, i.e. if there are $l$ ubiquitous items with a *support* of 100% than the number of baskets to consider is multiplied by a factor of $2^l$.
  Let us remark that we are also interested in the combinations with a single item and with no items since these are added to the baskets with no *ubiquitous* items referred to as *interesting baskets*.

For example if there are 4 *ubiquitous* items, the additional baskets containing at least one *ubiquitous* item is the following lattice, expressed as a collection of combinations, multiplied by the number of *interesting* baskets.

$$\{a, b, c, d\} \qquad = C\binom{4}{4} = \frac{4!}{4! \cdot 0!} = 1$$

$$\{a, b, c\} \; \{a, b, d\} \; \{a, c, d\} \; \{b, c, d\} \qquad = C\binom{4}{3} = \frac{4!}{3! \cdot 1!} = 4$$

$$\{a. b\} \; \{a, c\} \; \{a, d\} \; \{b, c\} \; \{b, d\} \; \{c, d\} \qquad = C\binom{4}{2} = \frac{4!}{2! \cdot 2!} = 6$$

$$\{a\} \; \{b\} \; \{c\} \; \{d\} \qquad = C\binom{4}{1} = \frac{4!}{1! \cdot 3!} = 4$$

*Figure 1: Combinations of 4 ubiquitous items*

This means that if we have $n$ different items, for each line, where a line $k$ represents a set of baskets with $k$ items. $b_k$, the expected number of baskets in line $k$ with above or equal a support of $s$ assuming an average support of $\bar{s}$ is:

*Equation 2*

$$b_k = c \binom{n}{k} \cdot \bar{s}^k$$

With the *ubiquitousness* parameter $u$, we separate the items into two sets U and S:
- $|U| - l$ items with *support* above $u$
- $|S| - m$ items with *support* below or equal $u$ but above the *support* $s$.

Since for items in $S$, $\bar{s}^k$ becomes very small as $k$ grows because of $s$ is below the threshold and much smaller than 1, the number of frequent baskets generated out of $S$ is much smaller than the total number of frequent baskets.

Assuming that all the items in $U$ have a support of 100%, and the number of *frequent* baskets, i.e. baskets created out of the total of *frequent* items including the items in $U$. If the number of frequent baskets out of the items in $S$ is $x$, then the number of *frequent* baskets including the *ubiquitous* items is:
- The *frequent* baskets with the different combinations of the items in $U$: $2^l \cdot x$
- Single *frequent* items (items in $S$) and the different combinations of the items in $U$, where at least one item out of $U$ has to exist to create a basket of at least 2 items: $(2^l - 1) \cdot m$
- All combinations of the items in $U$ with at least 2 items and no items out of $S$: $2^i - l - 1$

So the total number of *frequent* baskets is:

$$2^l \cdot x + (2^l - 1) \cdot m + 2^l - l - 1 \implies$$

*Equation 3*

$$(x + m + 1) \cdot 2^l - m - l - 1$$

After collecting all the *frequent* baskets, we need to check each basket and extract possible rules that satisfy the *confidence* threshold. Let us assume that a basket has on average $n$ items. This implies that $2^n - 2$ combinations (we exclude the cases where all items are either in the lhs or in the rhs) have to be checked against the confidence threshold.

This allows us to conclude that the *ubiquitous* items increase significantly the complexity of producing the rules out of the given *frequent* baskets as well.

Furthermore, let us investigate the quality of the generated rules while excluding the *ubiquitous* items. This will help us to measure the impact of removing these ubiquitous items. To do so, let us consider the lift parameter which is a measure comparing the likelihood of appearance of an item in a basket compared to the likelihood of item's appearance when picked randomly.

The effect of removing a *ubiquitous* item $U$ from a basket $B$ with items $\{b_1, b_2, \cdots, b_n\}$ is as follows:

$$\frac{lift(without\ Ubiquitous\ item)}{lift\ (with\ Ubiquitous\ item)} =$$

$$\frac{\frac{supprt\{B\}}{support\{b_1\}\cdot support\{b_2\}\cdots support\{b_n\}}}{\frac{support\{U,B\}}{support\{U\}\cdot support\{b_1\}\cdot support\{b_2\}\cdots support\{b_n\}}} = \frac{support\{U\}\cdot support\{B\}}{support\{U,B\}}$$

We can conclude that the more frequent the ubiquitous item, the less its effect, i.e., if the *ubiquitous* item has a frequency of 100% it has no effect at all on the lift. The less the *ubiquitous* items appear in the rule's basket the higher the lift of the rule without the *ubiquitous* items.

On the other hand, if the *ubiquitous* item appears in the rule's basket proportionately more than its *support,* then the rule without the *ubiquitous* item will have smaller *lift* than with it. So the *lift* is higher only if the *ubiquitous* item with the rest of the basket is more frequent than in all the transactions. However, since the item is *ubiquitous* we decide that it has little actual influence and thus this is less interesting.

Let's compare the rules with and without an *ubiquitous* item $U$. It is obvious that for all *frequent* baskets $\{U, B\}$ the baskets $\{B\}$ are also *frequent* given the Downward Closure Lemma. Therefore, let us investigate the impact on the rule's confidence.

Given the Downward Closure Lemma, we have that if a rule of the type $A \Rightarrow \{U, B\}$ is higher than the confidence threshold than the rule $A \Rightarrow B$ will be higher than the threshold as well. More formally, we have:

$Confidence = \frac{support\ (A,U,B)}{support\ (A)}$ and given the Downward Closure Lemma we have:

$$support\ (A,B) \geq support\ (A,U,B) \Rightarrow \frac{support\ (A,B)}{support\ (A)} \geq \frac{support\ (A,U,B)}{support\ (A)}$$

Thus the confidence is at least as high.

However when the ubiquitous item $U$ is in the left side of the rule there might be a case that the rule $A, U \Rightarrow B$ exists while the rule $A \Rightarrow B$ has a low *confidence*, i.e. $A, U \Rightarrow B$ has higher *confidence* than the rule $A \Rightarrow B$. This will only happen when there are not enough items $B$ in baskets containing $A$, and when adding $U$ to $A$, then fewer baskets containing $A$ but not containing $B$ (and obviously not containing $U$) are removed so that the rule still passes the *confidence* threshold. This does not indicate any relations between $A$ and $B$. On the contrary, since the confidence of $A \Rightarrow B$ is low, we have that $A$ actually does not indicate that $B$ will be in the basket. The predictor is in fact $U$. But since we defined that $U$ is not interesting because of its *ubiquitousness* property, then the rule $A, U \Rightarrow B$ is not interesting and it only implies $U \Rightarrow B$.

We can thus conclude that the quality of the rules are not compromised.

# 5. Experiment Results

In order to illustrate the impact and usefulness of the ubiquitousness parameter experiments have been performed experiments on both real-data sets as well as on artificially generated data sets using SAP HANA Predictive Analytics Library (PAL) (SAP SE, 2014) where the ubiquitous parameter is implemented in the Apriori algorithm. For each dataset, the running time, the number of items excluded, the number of generated rules have been measured in function of the ubiquitousness, the support and the confidence. The results are given in Table 1, Table 2 and Table 3.

At first, a real data set, that can be found at http://fimi.ua.ac.be/data/, has been used. This dataset contains anonymized traffic accident data (340,183 accidents). For more information, we refer to (Geurts, 以及其他人, 2003). The results for this dataset are given in Table 1 and represented in Figure 2.

|  | I | II | III | IV | V | VI | VII | VIII | IX |
|---|---|---|---|---|---|---|---|---|---|
| Ubiquitousness | 0.75 | 0.75 | 0.75 | 0.7 | 0.7 | 0.7 | 0.65 | 0.65 | 0.65 |
| Support | 0.6 | 0.5 | 0.4 | 0.5 | 0.4 | 0.3 | 0.5 | 0.4 | 0.3 |
| Confidence | 0.7 | 0.7 | 0.7 | 0.7 | 0.7 | 0.7 | 0.7 | 0.7 | 0.7 |
| Ignored Items | 27 | 27 | 27 | 31 | 31 | 31 | 40 | 40 | 40 |
| Created Rules | 1754 | 133830 | 5173056 | 7497 | 236882 | 10119511 | 56 | 421 | 13340 |
| Running Time | 3.73 | 46.49 | 491.69 | 8.96 | 76.28 | 897.88 | 1.35 | 3.63 | 13.77 |

*Table 1 Experiment 1*

Figure 2 displays the results of Table 1 with a logarithmic scale for a better representation for different values of the ubiquitousness parameter (0.75, 0.7 and 0.65). We can notice that the lower the threshold the lower the number of generated rules and thus the lower the computation time.

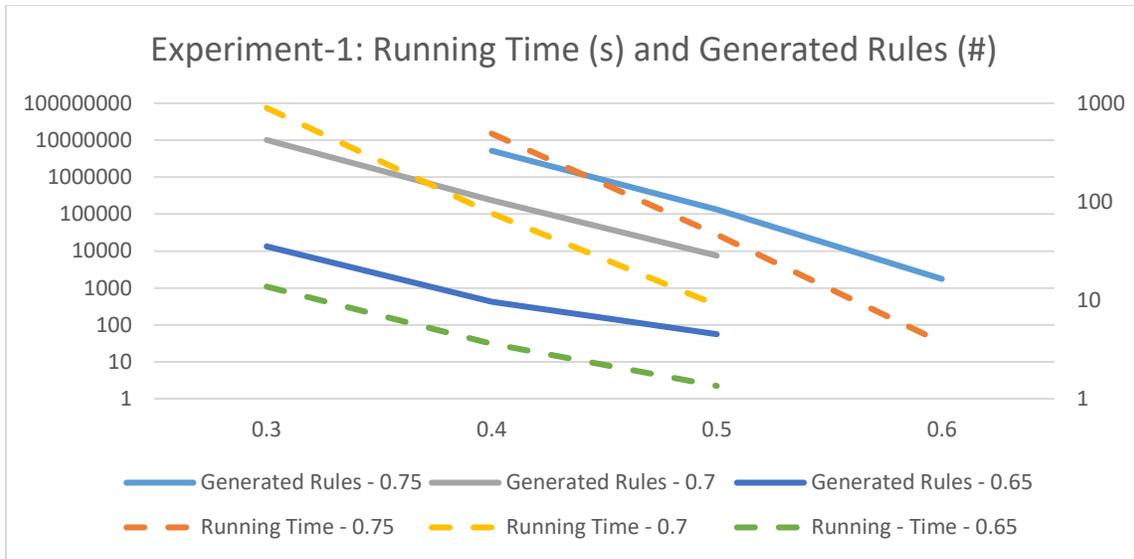

*Figure 2: Running Time and Generated Rules for Experiment-1 in function of* u *and* s.

In order to validate further our findings we performed to additional experiments for which we created a set of artificial datasets. The datasets were generated with a program that randomly adds items to a transaction ensuring that the items will appear at a given support. To ensure more control on the artificially generated dataset the generated items were not correlated thus requiring us to use low confidence.

In the second experiment, we created four datasets with the same items but we added step-by-step ubiquitous items, appearing in every transaction. We used the Apriori algorithm with support of 0.1 and confidence of 0.5. We show clearly that adding the ubiquitous items, in this case with zero entropy, has an exponential effect on the running time and number of created rules.

| Dataset | Items | Support | Rules | Time (ms.) |
|---|---|---|---|---|
| FIM1 | 10 | 0.3 | 62 | 53 |
|  | 5 | 0.5 |  |  |
|  | 0 | 1.0 |  |  |
| FIM2 | 10 | 0.3 | 1055 | 58 |
|  | 5 | 0.5 |  |  |
|  | 2 | 1.0 |  |  |
| FIM3 | 10 | 0.3 | 8395 | 83 |
|  | 5 | 0.5 |  |  |
|  | 4 | 1.0 |  |  |
| FIM4 | 5 | 0.5 | 33570 | 164 |
|  | 10 | 0.3 |  |  |
|  | 6 | 1.0 |  |  |

*Table 2: Experiment 2*

The results are given in Table 2 and Figure 3.

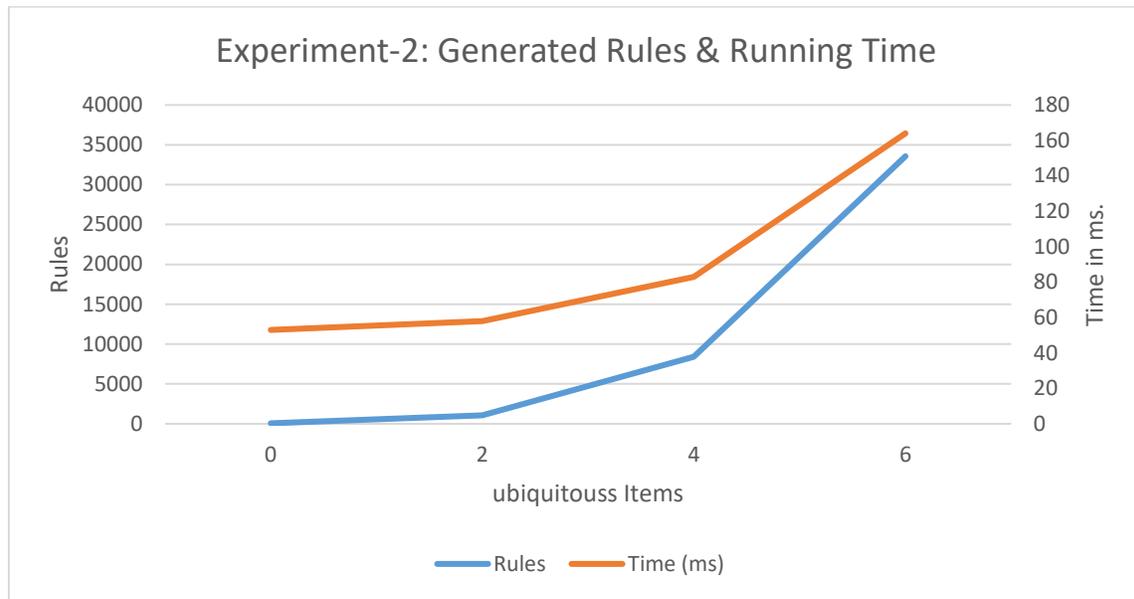

*Figure 3: Running Time and Generated Rules for Experiment-2.*

In experiment 3 we created a single database with fifteen regular items and six ubiquitous items with different support and show that with a high level of ubiquitousness, i.e. less ubiquitous items are filtered out, the performance changes exponentially both in matter of running time and number of created rules.

| FIM5 | I | II | III |
|---|---|---|---|
| **Ubiquitousness** | 0.7 | 0.85 | 0.95 |
| **Support** | 0.05 | 0.05 | 0.05 |
| **Confidence** | 0.05 | 0.05 | 0.05 |
| **Items with 0.3 support** | 10 | 10 | 10 |
| **Items with 0.5 support** | 5 | 5 | 5 |
| **Items with 0.8 support** | 2 | 2 | 2 |
| **Items with 0.9 support** | 2 | 2 | 2 |
| **Items with 1.0 support** | 2 | 2 | 2 |
| **Ignored Items** | 6 | 4 | 2 |
| **Created Rules** | 117 | 346 | 1014 |
| **Running Time (ms.)** | 182.9 | 205.461 | 256.885 |

*Table 3: Experiment 3*

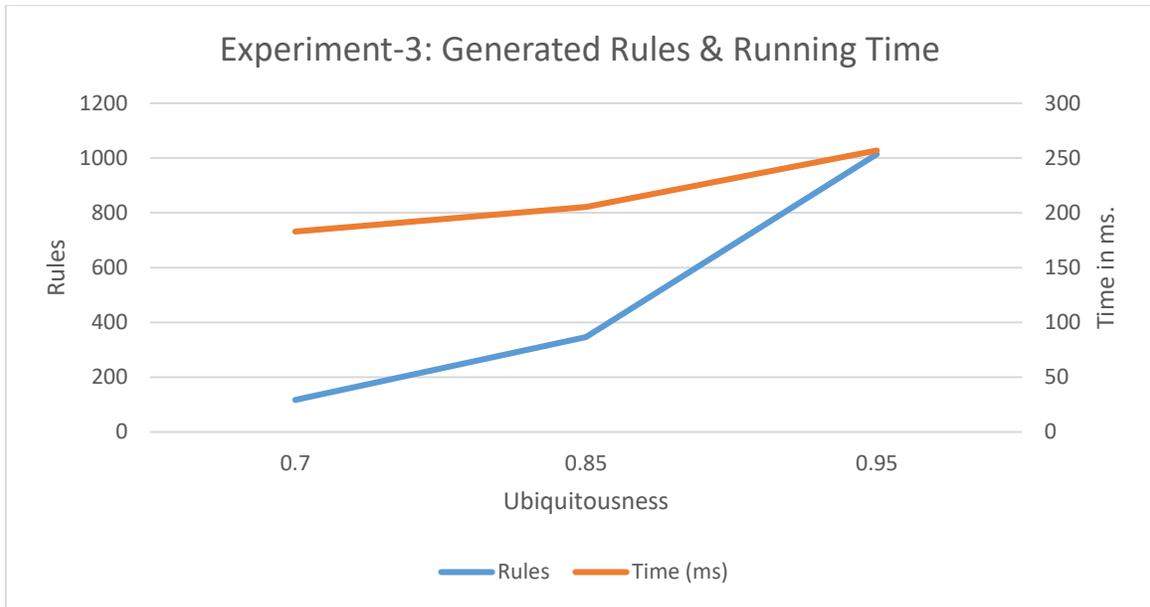

*Figure 4: Running Time and Generated Rules for Experiment-3.*

The results are given in Table 3 and Figure 4

# 6. Conclusion and Future Work

Frequent itemset mining tries to find relations between items based on their co-appearance in recorded transactions. To identify such relations between items, their co-occurrence in historical transactions are counted. However, the combinations of items that may appear in transactions increases rapidly as the number of items increases and it quickly becomes unfeasible to count them. This is why the FIM algorithms suggests an efficient method to eliminate specific items and item-combination. The algorithms assume that items that do not appear enough in the history dataset are not contributing to the understanding of the relations between items. This assumption is indeed natural but it ignores the fact that very frequent items do not contribute neither to the extraction of useful information about the items. As shown in this paper, frequent items are contributing significantly to the processing complexity of the algorithm such that they even prevent the algorithm from converging, and providing a result. In this paper, we suggested adding the so-called ubiquitousness parameter that eliminates items that are very frequent and so reducing the processing time significantly with minimum effect on the result quality.

The usage of the ubiquitousness parameter should be similar to the one of the support parameter. Both parameters should be fine-tuned to produce a consumable number of rules in an acceptable running time. The ubiquitousness parameter proved to have a greater effect both on the number of rules and on the processing time, while support usually filters out items that appear very infrequent. The exact values of the ubiquitousness parameter vary and are dependent on the dataset and the required ruleset size. trying to produce a histogram of item frequencies might help determining an initial value for those parameters.

The effect of items with similar entropy as item with low support, i.e. ubiquitous items is so far not extensively analyzed. We will further experiment how similar parameters can be used to increase the information gain while improving the performances.